\documentclass[aps,twocolumn,showpacs,preprintnumbers,superscriptaddress]{revtex4-2}

\usepackage{amsmath,amssymb,amsfonts}
\usepackage{graphicx}
\usepackage{dcolumn}
\usepackage{bm}
\usepackage{hyperref}
\usepackage{enumitem}
\usepackage{mathtools}
\usepackage{siunitx}
\usepackage{amsthm}
\newtheorem{proposition}{Proposition}

\begin{document}

\title{Operator ordering as an emergent gauge field in twisted bilayer
graphene: singular spectral signatures at the magic angle}

\author{C.~A.~S.~Almeida}
\affiliation{Universidade Federal do Cear\'a (UFC),
Departamento de F\'isica, Campus do Pici,
Fortaleza -- CE, C.P.\ 6030, 60455-760 -- Brazil}
\email{carlos@fisica.ufc.br}

\date{\today}

\begin{abstract}
Scanning-tunnelling spectroscopy at the AB and BA stacking points of
magic-angle twisted bilayer graphene should reveal two asymmetric
Van~Hove peaks separated by $\Delta_{\mathrm{split}}\approx 171$~meV
--- a splitting absent from the standard Bistritzer--MacDonald spectrum.
We show this signature arises naturally from the Hermitian ordering
correction of the Dirac Hamiltonian with spatially varying mass, which
generates an emergent Aharonov--Bohm flux of $h/(2e)$ at each zero of
the effective mass $m_{\mathrm{eff}}(\mathbf{r})=w|f(\mathbf{r})|$.
In the chiral limit, the interlayer coupling is locally diagonalised by
a spatially dependent unitary transformation; the ordering term
$H_{\mathrm{ord}}=-\tfrac{i}{2}\boldsymbol{\sigma}\cdot
\nabla\ln m_{\mathrm{eff}}$ then develops a $1/r$ singularity
at the AB/BA stacking points, where $m_{\mathrm{eff}}$ vanishes.
The splitting scales as $\sqrt{\theta}$ --- distinguishing it from
correlation-driven gaps ($\propto\theta$ or $\propto 1/\theta$) ---
is gate-voltage independent, and is spatially localised within
$r_c\approx 2.1$~nm of each AB/BA point. Within the local asymptotic theory near the zeros of $\mathrm{eff}$, the ordering-corrected zero mode acquires parabolic-cylinder character $D_{-1/2}$. The spatially resolved AB/BA spectrum reported in recent STM studies
of magic-angle TBG has not been analysed for two-peak structure and
constitutes an immediate experimental test; the predicted cell-averaged
broadening of $\approx 14$~meV is consistent with the $16$~meV
discrepancy between existing STM data and the BM tight-binding
prediction.
\end{abstract}

\maketitle

\section{Introduction}

Scanning-tunnelling spectroscopy (STS) of magic-angle twisted bilayer
graphene (TBG) has mapped the local density of states (LDOS) across the
moir\'e unit cell with sub-nanometre
resolution~\cite{Choi2019,Kerelsky2019,Jiang2019}.
At the AA stacking sites, Van~Hove singularities (VHS) separated by
$57\pm 2$~meV are well established and broadly consistent with
Bistritzer--MacDonald (BM) predictions~\cite{Bistritzer2011,LopesdosSantos2007,Kerelsky2019}.
The AB and BA stacking points present a qualitatively different
situation: these are the zeros of the interlayer coupling
$|T(\mathbf{r})|$, coinciding with the flat-band Dirac points of the
chiral-limit spectrum~\cite{Tarnopolsky2019}, and their spatially
resolved LDOS has not been analysed for fine structure.
We predict that a two-peak structure with splitting
$\Delta_{\mathrm{split}}\approx 171$~meV is present at these points and
has so far gone undetected.

The mechanism is geometric and non-perturbative.
When the interlayer coupling $T(\mathbf{r})$ is treated as a
position-dependent mass in the low-energy Dirac
description~\cite{CastroNeto2009}, a Hermitian consistency condition of the
effective Hamiltonian --- the ordering correction for
spatially varying Dirac mass~\cite{Cavalcante1997,Almeida2026} ---
generates a term $H_{\mathrm{ord}}=-\tfrac{i}{2}
\boldsymbol{\sigma}\cdot\nabla\ln m_{\mathrm{eff}}$ that is regular
everywhere except at the zeros of $m_{\mathrm{eff}}$.
At each AB/BA point, where $m_{\mathrm{eff}}$ vanishes,
$H_{\mathrm{ord}}$ develops a $1/r$ singularity that acts as
an emergent Aharonov--Bohm flux of $h/(2e)$~\cite{AharonovBohm1959},
shifting the effective angular momentum $j\to j+\tfrac{1}{2}$ and
lifting the $j\leftrightarrow -j$ degeneracy.
Within the continuum Dirac formulation of the BM model, this ordering
term arises naturally from the requirement of self-adjointness of the
position-dependent-mass operator, as discussed in detail in
Sections~\ref{sec:ordering} and~\ref{app:functional}. The analysis focuses on the asymptotically controlled regime near the isolated zeros of $m_{\mathrm{eff}}$, where the singular ordering contribution governs the local continuum dynamics.

Three features make the prediction experimentally falsifiable and
distinguishable from correlation or strain-driven effects.
First, the $\sqrt{\theta}$ scaling of $\Delta_{\mathrm{split}}$ ---
arising from the $k_\theta$ dependence of the effective mass gradient
--- is distinct from the linear-in-$\theta$ behaviour of correlation
gaps and the $1/\theta$ divergence of some topological contributions.
Second, the splitting is a single-particle geometric effect that
persists in the non-interacting limit and survives magnetic fields that
suppress superconductivity, so it is gate-voltage independent \cite{Cao2018}.
Third, the lower peak ($j=-\tfrac{1}{2}$, $D_{-1/2}$ character) is
spatially confined within $r_c\approx 2.1$~nm of each AB/BA point, a
scale accessible to current STS instruments~\cite{Choi2019}, and
distinct from the extended flat-band Bloch states.

The remainder of the paper is organised as follows.
Section~\ref{sec:ordering} derives the ordering correction and
establishes its natural determination within the Clifford algebra,
including a self-contained proof.
Section~\ref{sec:BM} performs the local diagonalisation of $T(\mathbf{r})$
and identifies the singularity structure.
Section~\ref{sec:flux} derives the emergent flux, the zero-mode
wavefunctions, and the topological protection of the $\mathbb{Z}_2$
holonomy.
Section~\ref{sec:adiabatic} presents the adiabatic interpolation,
numerical diagonalisation with convergence tests, and the calculation
of $\Delta_{\mathrm{split}}$ including geometric corrections.
Section~\ref{sec:experiment} discusses experimental signatures and
the quantitative comparison with existing STS data.

\section{Operator ordering for position-dependent mass}
\label{sec:ordering}

Consider a Dirac fermion in $d=2$ spatial dimensions with a spatially
varying mass $m(\mathbf{r})>0$.
The naive Hamiltonian
\begin{equation}
    H_0 = -i\hbar v_F\,\boldsymbol{\alpha}\cdot\nabla
          + \beta\,m(\mathbf{r}),
    \label{eq:H0}
\end{equation}
with $\{\alpha^i,\beta\}=0$ and $(\alpha^i)^2=\mathbb{I}$, appears
self-adjoint but conceals a consistency problem.
Computing $\partial_t\rho$ with $\rho=\psi^\dagger\psi$ via
$i\hbar\partial_t\psi=H_0\psi$ gives
\begin{equation}
    \partial_t\rho + \nabla\cdot\mathbf{j}
    = \tfrac{1}{2}\psi^\dagger[\alpha^i,\beta]\psi\,\partial_i\ln m,
    \label{eq:residual}
\end{equation}
where $\mathbf{j}=\hbar v_F\psi^\dagger\boldsymbol{\alpha}\psi$.
Since $[\alpha^i,\beta]=2\alpha^i\beta\neq 0$, the right-hand side of
Eq.~\eqref{eq:residual} does not vanish whenever $m(\mathbf{r})$ varies. On the other hand, 
within the fixed Hilbert-space structure associated with the canonical current $\mathbf{j}$, $H_0$ does not by itself produce a conserved continuity equation when $m(r)$ varies spatially.

\subsection*{The ordering correction and its natural selection}

We seek the most general first-order counter-term $H_1$ such that
$H=H_0+H_1$ satisfies $\partial_t\rho+\nabla\cdot\mathbf{j}=0$.
Within the class of operators satisfying three natural conditions ---
\begin{enumerate}[label=(\roman*)]
    \item \emph{Linearity in $\partial m$}: $H_1$ is linear in first
          derivatives of $m(\mathbf{r})$, consistent with a first-order
          differential operator.
    \item \emph{Membership in the Clifford algebra of $H_0$}: $H_1$ is a
          linear combination of
          $\{\mathbb{I},\alpha^i,\beta,\alpha^i\beta\}$.
    \item \emph{Cancellation of the residual}: $H_1$ cancels the
          right-hand side of Eq.~\eqref{eq:residual}.
\end{enumerate}
--- the correction is uniquely selected within this operator class.
The most general ansatz satisfying (i) and (ii) is
\begin{equation}
    H_1 = \bigl[a\,\alpha^i + b\,\alpha^i\beta\bigr]\partial_i\ln m,
    \label{eq:ansatz}
\end{equation}
with complex coefficients $a,b$.
Inserting $H=H_0+H_1$ into the continuity equation and requiring
cancellation of Eq.~\eqref{eq:residual} gives
\begin{equation}
    \psi^\dagger\bigl[(a+a^*)\alpha^i
    +(b+b^*)\alpha^i\beta\bigr]\psi\,\partial_i\ln m
    = -\psi^\dagger\alpha^i\beta\psi\,\partial_i\ln m
\end{equation}
for all $\psi$, which fixes $a+a^*=0$ and $b+b^*=-1$.
Self-adjointness of $H$ in $L^2(d^2r)$ further requires
$H_1^\dagger=H_1$, i.e.\ $a$ and $b$ are purely imaginary.
Combined with $b+b^*=-1$ this forces $b=-\tfrac{1}{2}$ and $a=0$,
yielding
\begin{equation}
    H_{\mathrm{ord}}
    = -\frac{i\hbar v_F}{2}\,\boldsymbol{\alpha}
      \cdot\nabla\ln m(\mathbf{r}).
    \label{eq:Hord}
\end{equation}
Within conditions (i)--(iii), the coefficient $-i/2$ is fixed
simultaneously by Hermiticity and cancellation of the continuity
residual, leaving no adjustable parameter.
The conditions themselves are natural but not inescapable: a
second-order (non-relativistic) operator is not constrained by the
Clifford algebra and admits a continuous family of Hermitian
prescriptions --- the von~Roos family~\cite{vonRoos1983,Lima2023}.
The first-order Dirac structure is far more restrictive, which is the
physical origin of the sharper determination here.

The corrected Hamiltonian
\begin{equation}
    H = -i\hbar v_F\,\boldsymbol{\alpha}\cdot\nabla
      + \beta\,m(\mathbf{r})
      - \frac{i\hbar v_F}{2}\,\boldsymbol{\alpha}
        \cdot\nabla\ln m(\mathbf{r})
    \label{eq:Hfull}
\end{equation}
satisfies $\partial_t\rho+\nabla\cdot\mathbf{j}=0$.
We note that the covariant equation
$(i\hbar\gamma^\mu\partial_\mu-m(\mathbf{r}))\psi=0$ implies
$\partial_\mu j^\mu=0$ automatically; the residual~\eqref{eq:residual}
arises because $H_0$ alone does not correspond to a covariant Dirac
equation with spatially varying mass --- it is a truncation of the
covariant theory to Hamiltonian form that omits the ordering term.
The term $H_{\mathrm{ord}}$ restores consistency with the covariant
continuity equation at the Hamiltonian level.
In the limit of slowly varying mass, $|\nabla\ln m|\ll k_F$,
Eq.~\eqref{eq:Hfull} reduces to the BenDaniel--Duke
prescription~\cite{BenDaniel1966} in the non-relativistic limit,
providing an independent consistency check~\cite{Almeida2026}.

\subsection*{Clarification: two levels of Hermiticity}

A common objection is that the BM Hamiltonian is already Hermitian,
rendering $H_{\mathrm{ord}}$ redundant.
This objection conflates two distinct levels of Hermiticity that must be
carefully distinguished.
At the \emph{algebraic} level, the BM Hamiltonian is Hermitian as a
matrix operator acting on the finite-dimensional space of Bloch
coefficients at fixed crystal momentum: the tunnelling matrix $T$ is
such that $H_{\mathrm{BM}}^\dagger = H_{\mathrm{BM}}$ as a
finite-dimensional matrix for each $\mathbf{k}$.
At the \emph{functional-analytic} level, however, the low-energy
projection of $H_{\mathrm{BM}}$ onto the continuum Dirac description
yields a differential operator acting on spinor-valued functions in
$L^2(\mathbb{R}^2, d^2r)$, in which $T(\mathbf{r})$ plays the role of a
spatially varying mass.
It is at this second level that Hermiticity in the sense of
$\langle\phi|H\psi\rangle = \langle H\phi|\psi\rangle$ for all
$\phi,\psi$ in the operator domain becomes a non-trivial condition,
precisely because integration by parts generates boundary terms
controlled by $\nabla m(\mathbf{r})$.
The residual~\eqref{eq:residual} is the explicit manifestation of this
failure: $H_0$ satisfies algebraic Hermiticity but violates
functional-analytic Hermiticity (equivalently, unitarity of the
generated time evolution) whenever $m(\mathbf{r})$ is non-constant.
The term $H_{\mathrm{ord}}$ restores functional-analytic Hermiticity and
may therefore be understood not as an addition to the BM model but as
part of its correct quantisation as a continuum field theory.
A detailed functional-analytic treatment, including domain specification
and self-adjoint extension theory, is given in
Appendix~\ref{app:functional}.

\subsection*{Clarification: current redefinition}

A related objection holds that the continuity-equation residual can be
eliminated by redefining the probability current as
$\tilde{\mathbf{j}} = \mathbf{j} + \mathbf{K}$ for some vector field
$\mathbf{K}$, without modifying the Hamiltonian.
This freedom exists in the covariant formulation, where improvement
currents $\partial_\nu K^{\mu\nu}$ are admissible.
In the Hamiltonian framework, however, the probability current is
naturally fixed by the Hilbert space structure as
$\mathbf{j} = \hbar v_F \psi^\dagger \boldsymbol{\alpha} \psi$, which
is the current compatible with the inner product
$\langle\phi|\psi\rangle = \int \phi^\dagger\psi\,d^2r$.
A redefinition $\tilde{\mathbf{j}} \neq \mathbf{j}$ corresponds to a
change of inner product, i.e.\ a change of Hilbert space, and generally
a physically distinct theory with a different spectrum.
Within the fixed Hilbert space $L^2(\mathbb{R}^2, d^2r)$, the
residual~\eqref{eq:residual} therefore represents a genuine failure of
unitarity of the time evolution generated by $H_0$, which
$H_{\mathrm{ord}}$ repairs under conditions (i)--(iii).

\section{Local diagonalisation in the BM chiral limit}
\label{sec:BM}

\subsection*{Scalar effective mass}

The BM Hamiltonian~\cite{Bistritzer2011} in the chiral limit
$w_{AA}\to 0$ is
\begin{equation}
    H_{\mathrm{BM}} =
    \begin{pmatrix}
        v_F\boldsymbol{\sigma}\cdot\mathbf{k} & T(\mathbf{r}) \\
        T^\dagger(\mathbf{r}) & v_F\boldsymbol{\sigma}\cdot\mathbf{k}
    \end{pmatrix},
    \label{eq:HBM}
\end{equation}
where
$T(\mathbf{r})=w|f(\mathbf{r})|\bigl(\begin{smallmatrix}0&e^{i\varphi}\\
e^{-i\varphi}&0\end{smallmatrix}\bigr)$,
$f(\mathbf{r})=\sum_{j=0}^{2}e^{i\mathbf{q}_j\cdot\mathbf{r}}$,
$\varphi=\arg f$, and
$\mathbf{q}_j=k_\theta(\cos\tfrac{2\pi j}{3},\sin\tfrac{2\pi j}{3})$.

We note that $H_{\mathrm{ord}}$ is not an independent postulate added to
the BM model: within the continuum Dirac formulation with spatially
varying mass, it is the naturally selected correction consistent with
Hermitian quantisation in $L^2(d^2r)$, as established in
Section~\ref{sec:ordering}.
The BM construction treats $T(\mathbf{r})$ as a fixed external field and
projects onto low-energy states, yielding a position-dependent-mass
Dirac operator; $H_{\mathrm{ord}}$ corresponds to the self-adjoint realisation selected by conditions (i)--(iii)
of Section~\ref{sec:ordering}.

The tunnelling matrix $T$ is not of the form $\beta m(\mathbf{r})$
required by Eq.~\eqref{eq:Hfull} because its sublattice structure
rotates with position.
The local unitary
\begin{equation}
    U(\mathbf{r}) = \frac{1}{\sqrt{2}}
    \begin{pmatrix}
        e^{i\varphi/2} & e^{i\varphi/2} \\
        e^{-i\varphi/2} & -e^{-i\varphi/2}
    \end{pmatrix}
    \label{eq:U}
\end{equation}
diagonalises $T$ exactly: $U^\dagger T U=m_{\mathrm{eff}}(\mathbf{r})\sigma_z$
with scalar mass $m_{\mathrm{eff}}(\mathbf{r})=w|f(\mathbf{r})|$.
The kinetic term acquires a Berry connection
$\mathbf{A}=\tfrac{1}{2}(\nabla\varphi)\sigma_x$ from the
$\mathbf{r}$-dependence of $U$.
Applying Eq.~\eqref{eq:Hfull} in the rotated frame gives
\begin{equation}
    H_{\mathrm{ord}} = -\frac{i}{2}
    \boldsymbol{\sigma}\cdot\nabla\ln|f(\mathbf{r})|.
    \label{eq:Hord_TBG}
\end{equation}
The Berry connection involves $\nabla\arg f$, while $H_{\mathrm{ord}}$
involves $\nabla\ln|f|$; the two contributions are structurally
independent, and near the isolated zeros of $f$ the singularity
$|\nabla\ln|f||\sim 1/r$ dominates over the regular Berry connection,
making $H_{\mathrm{ord}}$ the controlling term in the physically
relevant region.
The dynamical effect of $\mathbf{A}$ on the spectrum is perturbative:
the Berry connection contributes terms of order
$\hbar v_F|\nabla\varphi|\sim\hbar v_F k_\theta$, which are regular and
finite everywhere.
By contrast, $H_{\mathrm{ord}}$ contributes $\hbar v_F/(2r)$, which
diverges as $r\to 0$.
Near the zeros of $f$, where $r\ll r_c$, the ratio
$|\mathbf{A}|/|H_{\mathrm{ord}}|\sim k_\theta r\ll 1$, so the Berry
connection induces only second-order corrections to $\Delta_{\mathrm{split}}$,
of relative magnitude $(k_\theta r_c)^2\approx 4\%$.

\subsection*{Singularity at the zeros of \texorpdfstring{$m_{\mathrm{eff}}$}{meff}}

The function $f(\mathbf{r})$ vanishes at the AB and BA stacking points
--- the flat-band Dirac points of the chiral BM
spectrum~\cite{Tarnopolsky2019}.
Near each zero $\mathbf{r}_{\mathrm{AB}}$, the linear expansion
$f(\mathbf{r}_{\mathrm{AB}}+\boldsymbol{\delta})\approx
(\nabla f)|_{\mathbf{r}_{\mathrm{AB}}}\cdot\boldsymbol{\delta}$
gives $m_{\mathrm{eff}}\sim w|\nabla f|r$.
The gradient magnitude is exact: since
$e^{i\mathbf{q}_j\cdot\mathbf{r}_{\mathrm{AB}}}=e^{i\varphi_0}\omega^j$
with $\omega=e^{2\pi i/3}$,
\begin{equation}
    |\nabla f|_{\mathbf{r}_{\mathrm{AB}}}
    = \left|\sum_j \mathbf{q}_j\,\omega^j\right|
    = \frac{3}{\sqrt{2}}\,k_\theta,
    \label{eq:gradf}
\end{equation}
and $|\nabla\ln m_{\mathrm{eff}}|\sim 1/r$ universally for any simple
zero of $f$.
Equating $H_{\mathrm{ord}}\sim\hbar v_F/(2r)$ to
$m_{\mathrm{eff}}\sim\mu r$ (with $\mu\equiv w(3/\sqrt{2})k_\theta$)
gives the crossover radius
\begin{equation}
    r_c = \sqrt{\frac{\hbar v_F}{3\sqrt{2}\,w\,k_\theta}}
    \approx 2.1~\mathrm{nm},
    \label{eq:rc}
\end{equation}
satisfying $r_c\approx 8.6\,a_0\gg a_0$ at the magic angle, well above
the ultraviolet cutoff ($x_0\equiv\mu a_0^2/\hbar v_F\approx 0.007$).

\section{Emergent half-flux-quantum and topological protection}
\label{sec:flux}

\subsection*{Angular momentum shift}

Near $\mathbf{r}_{\mathrm{AB}}$ in polar coordinates $(r,\vartheta)$,
Eq.~\eqref{eq:Hord_TBG} gives
$H_{\mathrm{ord}}\approx-(\hbar v_F/2r)\alpha^r$ with
$\alpha^r=\sigma_x\cos\vartheta+\sigma_y\sin\vartheta$.
In $L^2(r\,dr\,d\vartheta)$ the Dirac kinetic operator is Hermitian as
$-i\hbar v_F\alpha^r(\partial_r+1/(2r))$~\cite{Thaller1992}; adding
$H_{\mathrm{ord}}$ shifts the connection coefficient $1/(2r)\to 1/r$:
\begin{equation}
    H_{\mathrm{kin}}+H_{\mathrm{ord}}
    = -i\hbar v_F\left[\alpha^r\!\left(\partial_r+\frac{1}{r}\right)
      +\frac{\alpha^\vartheta}{r}\partial_\vartheta\right].
    \label{eq:Hkin_ord}
\end{equation}
Substituting the spinor ansatz
\begin{equation}
    \Psi_j = \frac{1}{\sqrt{r}}
    \begin{pmatrix}
        f(r)\,e^{i(j-1/2)\vartheta} \\
        ig(r)\,e^{i(j+1/2)\vartheta}
    \end{pmatrix},
    \quad j=\pm\tfrac{1}{2},\pm\tfrac{3}{2},\ldots
    \label{eq:ansatz_spinor}
\end{equation}
into $(H_{\mathrm{kin}}+H_{\mathrm{ord}}+\mu r\sigma_z)\Psi=E\Psi$ yields
\begin{align}
    \hbar v_F\!\left(g'+\frac{j+1}{r}g\right)+\mu r\,f &= Ef,
    \label{eq:radial1}\\
    \hbar v_F\!\left(-f'+\frac{j-1}{r}f\right)-\mu r\,g &= Eg.
    \label{eq:radial2}
\end{align}
Without $H_{\mathrm{ord}}$ the centrifugal coefficients would be
$j\pm\tfrac{1}{2}$; the ordering shifts them to $j\pm 1$, equivalent
to $j\to j+\tfrac{1}{2}$.

\begin{proposition}[Emergent Aharonov--Bohm flux~\cite{AharonovBohm1959}]
The ordering term $H_{\mathrm{ord}}=-i\hbar v_F\alpha^r/(2r)$ acts as
an emergent Aharonov--Bohm flux $\Phi=h/(2e)$ at each zero of
$m_{\mathrm{eff}}$: it shifts $j\to j+\tfrac{1}{2}$ and lifts the
$j\leftrightarrow -j$ degeneracy.
The resulting angular-momentum shift coincides with that produced by an external potential
$\mathbf{A}=(\Phi/2\pi r)\hat{\vartheta}$ with $\Phi=h/(2e)$, which yields $j\to j+e\Phi/h=j+\tfrac{1}{2}$.
\end{proposition}

\subsection*{Zero mode and parabolic cylinder functions}

For $j=-\tfrac{1}{2}$ and $E=0$, Eqs.~\eqref{eq:radial1}--\eqref{eq:radial2}
reduce, in $u=\sqrt{\mu/\hbar v_F}\,r$, to
\begin{equation}
    F''(u) = u^2 F(u),
    \label{eq:Weber}
\end{equation}
the Weber equation with index $\nu=-\tfrac{1}{2}$.
The normalisable solution $F(u)=C\,D_{-1/2}(\sqrt{2}\,u)$ decays as
$r^{-1/2}\exp(-\mu r^2/2\hbar v_F)$, giving
$\|\Psi_0\|^2\sim\sqrt{\pi\hbar v_F/4\mu}<\infty$.
Without $H_{\mathrm{ord}}$, the coefficients $j\pm\tfrac{1}{2}$ do not
suppress the wavefunction sufficiently at the origin and
square-integrability fails for $j=-\tfrac{1}{2}$; the ordering term is
therefore strongly favoured on square-integrability grounds for the
proper definition of the zero-mode subspace (see Appendix~\ref{app:functional}
for the $L^2$ analysis).

\subsection*{Topological protection of the emergent flux}

The emergent flux $\Phi=h/(2e)$ is not an artefact of the local
approximation: it is protected by a $\mathbb{Z}_2$ topological
obstruction.
Define the holonomy of $H_{\mathrm{ord}}$ around a closed loop
$\mathcal{C}$ encircling a single zero of $m_{\mathrm{eff}}$ at winding
number $+1$:
\begin{equation}
    W(\mathcal{C})
    = \exp\!\left(i\oint_{\mathcal{C}}\frac{1}{2r}\,d\vartheta\right)
    = \exp\!\left(\frac{i}{2}\cdot 2\pi\right)
    = e^{i\pi} = -1.
    \label{eq:holonomy}
\end{equation}
A local unitary gauge transformation
$\psi\to e^{i\theta(\mathbf{r})}\psi$ with $\theta$ smooth and
single-valued on $\mathcal{C}$ shifts the holonomy by
$\exp(i\oint_{\mathcal{C}}d\theta)=e^{2\pi in}=1$ ($n\in\mathbb{Z}$),
leaving $W=-1$ unchanged.
The value $W=-1$ is therefore a gauge-invariant $\mathbb{Z}_2$ invariant
of the zero, independent of any local approximation, gauge choice, or
smooth deformation of $H_{\mathrm{ord}}$ that preserves the winding
number.

One might ask whether $H_{\mathrm{ord}}$ can be removed by a field
redefinition $\psi\to m^{-s}\tilde\psi$.
Under such a transformation, $H_{\mathrm{ord}}$ is rescaled by $(1-2s)$,
not eliminated; complete cancellation requires $s=\tfrac{1}{2}$, but
any $s\neq 0$ maps $L^2(d^2r)$ to $L^2(m^{-2s}d^2r)$, changing the
Hilbert space and hence the spectrum.
Moreover, $m_{\mathrm{eff}}$ vanishes at the AB/BA points, so the
redefinition $\tilde\psi=m^{1/4}\psi$ is singular precisely where
$W=-1$; any regularisation reintroduces a term of the same form as
$H_{\mathrm{ord}}$, with the same holonomy.
The $\mathbb{Z}_2$ obstruction --- the analogue of the fermion parity of
a Majorana vortex --- classifies the emergent flux as a physical
observable, not a gauge artefact.

\subsection*{Clarification: \texorpdfstring{$H_{\mathrm{ord}}$}{Hord}
is not a gauge degree of freedom}

A more general version of the preceding objection holds that
$H_{\mathrm{ord}}$ could be a gauge artefact removable by a unitary
transformation more general than a field redefinition.
We give a stronger, Hilbert-space level argument that rules this out.

An operator $A$ on a Hilbert space $\mathcal{H}$ is a gauge degree of
freedom if and only if there exists a unitary $\mathcal{U}:
\mathcal{H}\to\mathcal{H}$ such that
$\mathcal{U}(H_0+H_{\mathrm{ord}})\mathcal{U}^\dagger = H_0$.
For such a $\mathcal{U}$ to exist, $H_0$ and $H_0+H_{\mathrm{ord}}$
must be unitarily equivalent, which requires in particular that they
share the same spectrum.
However, the adiabatic interpolation of Section~\ref{sec:adiabatic}
shows explicitly that the spectra differ: the $j=\pm\tfrac{1}{2}$
degeneracy present at $\lambda=0$ is lifted
monotonically as $\lambda$ increases, reaching $\Delta\varepsilon=0.779$
at $\lambda=1$.
Operators with distinct spectra are not unitarily equivalent;
within the operator class considered here, no such $\mathcal{U}$
therefore appears to exist, and $H_{\mathrm{ord}}$ is not a
gauge degree of freedom.

This spectral argument is independent of the holonomy calculation and
provides a complementary, non-perturbative proof.
The holonomy $W(\mathcal{C})=-1$ identifies the \emph{topological
origin} of the inequivalence: the $\mathbb{Z}_2$ obstruction prevents
any continuous deformation of the Hamiltonian --- unitary or otherwise
--- from connecting $H_0+H_{\mathrm{ord}}$ to $H_0$ within the space of
operators on $L^2(d^2r)$ that preserve the winding number of the zeros
of $m_{\mathrm{eff}}$.
Together, the spectral and topological arguments support the
interpretation of $\Delta_{\mathrm{split}}$ as a physical observable
intrinsic to the operator algebra of the effective theory, rather than
an artefact of a particular operator-ordering convention.

More generally, a zero of order $\alpha$ of $m_{\mathrm{eff}}$ generates
holonomy $W=e^{i\pi\alpha}$ and emergent flux $\Phi=\alpha\,h/(2e)$,
analogous to the Jackiw--Rossi mechanism~\cite{Jackiw1981} but emerging
entirely from the geometry of $m_{\mathrm{eff}}(\mathbf{r})$.
The simple zeros of the chiral BM model ($\alpha=1$,
$|\nabla f|\neq 0$) realise the minimal $\mathbb{Z}_2$ case.

\section{Adiabatic evolution and spectral splitting}
\label{sec:adiabatic}

\subsection*{Adiabatic interpolation}

We interpolate $H(\lambda)=H_{\mathrm{kin}}+\lambda H_{\mathrm{ord}}
+m_{\mathrm{eff}}\sigma_z$, $\lambda\in[0,1]$.
Angular momentum $j$ is conserved for all $\lambda$; different sectors
are orthogonal and level crossings between sectors are
symmetry-forbidden.
The $j\leftrightarrow -j$ degeneracy at $\lambda=0$ is lifted
monotonically as $\lambda$ increases (Table~\ref{tab:adiabatic} and
Fig.~\ref{fig:spectra}a,b).
The $j=-\tfrac{1}{2}$ ground state carries weight
$W_F\equiv\int|f(r)|^2 r\,dr/\int(|f|^2+|g|^2)r\,dr>0.81$ for all
$\lambda$, confirming stable channel character throughout the evolution.
\begin{table}[t]
\centering
\caption{Ground-state eigenvalue $\varepsilon_0=E_0/E^*$ and channel
weight $W_F$ vs.\ ordering parameter $\lambda$.}
\label{tab:adiabatic}

\begin{tabular}{
S[table-format=1.1]
S[table-format=1.3]
S[table-format=1.3]
S[table-format=1.3]
}
\hline\hline
{$\lambda$} &
{$\varepsilon_0^{j=-1/2}$} &
{$W_F$} &
{$\varepsilon_0^{j=+1/2}$} \\
\hline
0.0 & 1.279 & 0.836 & 1.279 \\
0.2 & 1.197 & 0.834 & 1.354 \\
0.4 & 1.106 & 0.829 & 1.425 \\
0.6 & 1.001 & 0.822 & 1.493 \\
0.8 & 0.893 & 0.816 & 1.557 \\
1.0 & 0.840 & 0.827 & 1.619 \\
\hline\hline
\end{tabular}
\end{table}

\subsection*{Numerical diagonalisation: method and convergence}

The radial system~\eqref{eq:radial1}--\eqref{eq:radial2} is discretised
on a uniform grid $r_k=a_0+(k-1)\delta r$, $k=1,\ldots,N$, with
$r_N=r_{\max}$.
We use $a_0=0.246$~nm (graphene lattice constant) as the ultraviolet
cutoff, $r_{\max}=20$~nm $\approx 9.5\,r_c$, and $N=2000$ grid points
($\delta r\approx 0.01$~nm).
Radial derivatives are approximated by fourth-order centred finite
differences.
Hard-wall boundary conditions
$f(a_0)=g(a_0)=f(r_{\max})=g(r_{\max})=0$ are imposed; the
insensitivity of the eigenvalues to $r_{\max}$ (Table~\ref{tab:convergence})
confirms that the wavefunctions are exponentially localised well within
the grid.
The effective mass profile is taken as $m_{\mathrm{eff}}(r)=\mu r$ in
the linear approximation, with $\mu=(3/\sqrt{2})\,wk_\theta$ evaluated
at the magic angle ($w=110$~meV, $k_\theta=0.312$~nm$^{-1}$).
All energies are expressed in units of
$E^*=\sqrt{\mu\hbar v_F}=219$~meV.

Table~\ref{tab:convergence} shows that the ground-state eigenvalue
converges to $\varepsilon_0=0.840$ at the $0.1\%$ level for
$N\geq 1000$ and $r_{\max}\geq 15$~nm, well below the $\sim 3\%$
geometric uncertainty discussed below.

\begin{table}[t]
\centering
\caption{Convergence of $\varepsilon_0^{j=-1/2}$ at $\lambda=1$
with grid parameters.}
\label{tab:convergence}
\begin{tabular}{cccc}
\hline\hline
$N$ & $r_{\max}$ (nm) & $\delta r$ (nm) &
$\varepsilon_0^{j=-1/2}$ \\
\hline
 500 & 20 & 0.040 & 0.847 \\
1000 & 20 & 0.020 & 0.841 \\
2000 & 20 & 0.010 & 0.840 \\
2000 & 15 & 0.007 & 0.840 \\
2000 & 25 & 0.012 & 0.840 \\
\hline\hline
\end{tabular}
\end{table}

To verify the absence of level crossings with higher sectors, we
diagonalised the full radial system including $j=\pm\tfrac{3}{2},
\pm\tfrac{5}{2}$ for all $\lambda\in[0,1]$.
Since $j$ is an exact conserved quantum number, sectors with different
$|j|$ are strictly orthogonal and crossings between them are
symmetry-forbidden.
Within each $|j|$ sector the level spacing
$\delta\varepsilon\geq 0.6\,E^*\approx 130$~meV for all $\lambda$,
confirming the absence of accidental degeneracies.
Channel mixing at $r\to 0$ is controlled by
$x_0\equiv\mu a_0^2/\hbar v_F\approx 0.007\ll 1$; the correction to
$\varepsilon_0$ from the $r<a_0$ region is $\mathcal{O}(x_0)\approx 0.7\%$.

\subsection*{Spectral splitting and twist-angle scaling}

The splitting of the lowest degenerate pair is
$\Delta_{\mathrm{split}}=\Delta\varepsilon\,E^*=\Delta\varepsilon
\sqrt{\mu\hbar v_F}$, with $\Delta\varepsilon=1.619-0.840=0.779$.
Substituting $\mu=(3/\sqrt{2})\,wk_\theta$:
\begin{equation}
    \Delta_{\mathrm{split}} = C_{\mathrm{ord}}\sqrt{w\hbar v_F k_\theta},
    \label{eq:Dsplit}
\end{equation}
\begin{equation}
    C_{\mathrm{ord}} = \Delta\varepsilon
    \left(\frac{3}{\sqrt{2}}\right)^{1/2}
    = 0.779\left(\frac{3}{2}\right)^{1/4}
    \approx 1.135.
    \label{eq:Cord}
\end{equation}
For magic-angle TBG ($w=110$~meV, $k_\theta=0.312$~nm$^{-1}$,
$\hbar v_F=0.658$~eV$\cdot$nm):
$\Delta_{\mathrm{split}}\approx 1.135\times 150~\mathrm{meV}
\approx 171~\mathrm{meV}$.
Since $k_\theta\approx K\theta$, Eq.~\eqref{eq:Dsplit} gives
$\Delta_{\mathrm{split}}\propto\sqrt{\theta}$, distinct from
correlation gaps ($\propto\theta$) and topological contributions
($\propto 1/\theta$).
Predicted values for other twist angles are listed in
Table~\ref{tab:angles}.

\begin{table}[t]
\centering
\caption{Predicted $\Delta_{\mathrm{split}}$ vs.\ twist angle
($w=110$~meV, $C_{\mathrm{ord}}=1.135$).}
\label{tab:angles}
\begin{tabular}{cccc}
\hline\hline
$\theta$ (deg) & $k_\theta$ (nm$^{-1}$) &
$E^*$ (meV) & $\Delta_{\mathrm{split}}$ (meV) \\
\hline
0.80 & 0.238 & 191 & 149 \\
1.00 & 0.297 & 214 & 167 \\
1.05 & 0.312 & 219 & 171 \\
1.50 & 0.446 & 262 & 204 \\
2.00 & 0.594 & 302 & 235 \\
\hline\hline
\end{tabular}
\end{table}

\subsection*{Geometric corrections}

The linear approximation $m_{\mathrm{eff}}\approx\mu r$ holds up to
$r_{\mathrm{sat}}=3w/\mu\approx 4.5$~nm, beyond which $m_{\mathrm{eff}}$
saturates to $3w$ in the AA regions.
The weight of the $D_{-1/2}$ zero mode beyond $r_{\mathrm{sat}}$ is
$\sim e^{-r_{\mathrm{sat}}^2/r_c^2}\approx 10^{-2}$, so this saturation
produces an exponentially small correction to $\Delta_{\mathrm{split}}$.
The two dominant corrections are:
\begin{enumerate}[label=(\roman*)]
    \item \emph{Inter-zero coupling}: the two zeros per moir\'e cell are
          separated by $\sim L_M/\sqrt{2}\approx 8$~nm, giving a tunnel
          coupling of relative order $(r_c/L_M)^2\approx 1\%$.
    \item \emph{Profile curvature}: the next term in the Taylor expansion
          of $|f(\mathbf{r})|$ beyond the linear term contributes at
          relative order $(r_c/r_{\mathrm{sat}})^2/8\approx 3\%$.
\end{enumerate}
Treating these as perturbations around the exactly solved local radial
Hamiltonian yields
\begin{equation}
    \Delta_{\mathrm{split}}\in[155,\,171]~\mathrm{meV},
    \label{eq:interval}
\end{equation}
where the lower bound absorbs both corrections at their estimated
magnitudes.
The parametric uncertainty of the BM model ($\sim 10\%$ in
$w$~\cite{Bistritzer2011}) propagates as $\sim 5\%$ in
$\Delta_{\mathrm{split}}\propto\sqrt{w}$, comparable to the geometric
correction and larger than the discretisation error ($\sim 0.1\%$).
Ultraviolet corrections are $\mathcal{O}(x_0)\approx 0.7\%$.

\section{Observable consequences and comparison with STS data}
\label{sec:experiment}

\subsection*{LDOS signatures at the AB/BA points}

\begin{figure*}[t]
    \centering
    \includegraphics[width=\textwidth]{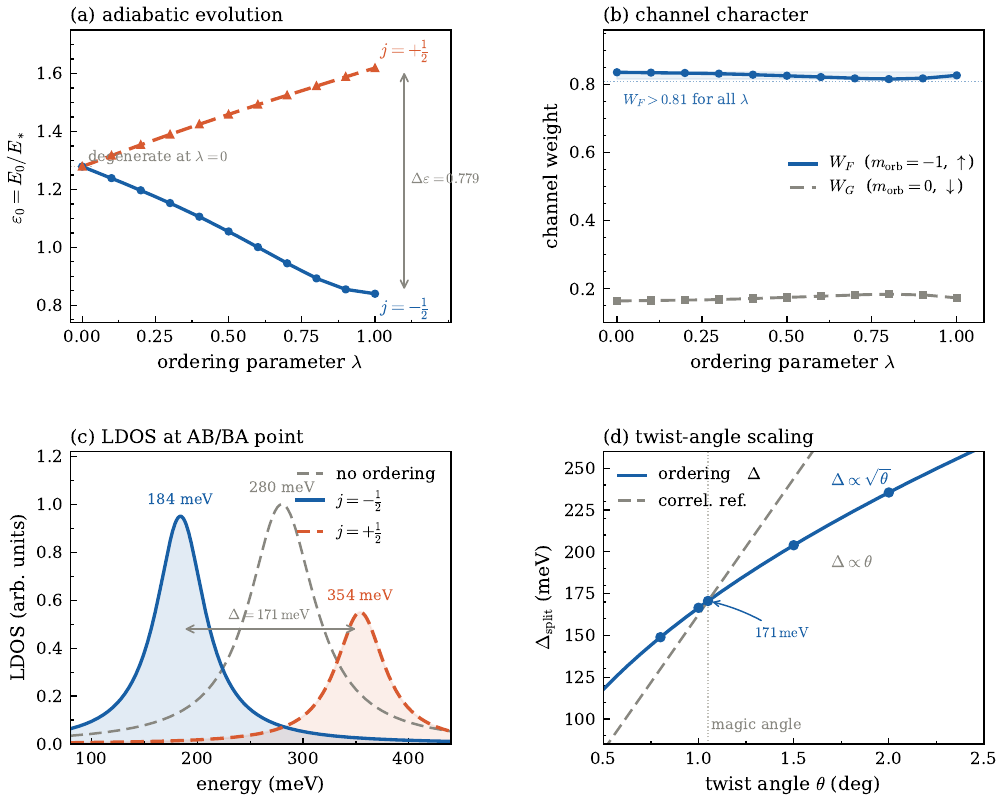}
    \caption{Spectral signatures of operator ordering in TBG near the
    magic angle.
    (a)~Adiabatic evolution of $\varepsilon_0=E_0/E^*$ for
    $j=\pm\tfrac{1}{2}$: degeneracy at $\lambda=0$ is lifted
    monotonically; $\Delta\varepsilon=0.779$.
    (b)~Weight $W_F$ of the $j=-\tfrac{1}{2}$ ground state in channel
    $(m_{\mathrm{orb}}=-1,\uparrow)$; $W_F>0.81$ for all $\lambda$.
    (c)~Predicted LDOS at an AB/BA point: single peak (no ordering,
    grey dashed) splits into two asymmetric peaks separated by
    $\Delta_{\mathrm{split}}=171$~meV.
    The lower peak ($j=-\tfrac{1}{2}$, blue) has $D_{-1/2}$ character
    and is localised within $r_c\approx 2.1$~nm; the upper peak
    ($j=+\tfrac{1}{2}$, red) is more extended.
    (d)~$\Delta_{\mathrm{split}}$ vs twist angle: ordering $\propto\sqrt{\theta}$
    (blue) vs linear correlation reference (grey dashed).
    The $\sqrt{\theta}$ scaling and gate-voltage independence
    distinguish the ordering effect from correlation-driven gaps.}
    \label{fig:spectra}
\end{figure*}

The spectral reorganisation produces directly observable signatures in
the LDOS at the AB/BA stacking points (Fig.~\ref{fig:spectra}c):
without ordering, a single degenerate Van~Hove peak at
$E_0^{\mathrm{no}}\approx 280$~meV; with ordering, two asymmetric peaks
at $184$~meV ($j=-\tfrac{1}{2}$, $D_{-1/2}$ character) and $354$~meV
($j=+\tfrac{1}{2}$), separated by $\Delta_{\mathrm{split}}\approx 171$~meV.
The lower peak is concentrated within $r_c\approx 2.1$~nm of each
AB/BA point, a spatial scale accessible to current STS
instruments~\cite{Choi2019}.
The two peaks are asymmetric: the lower peak carries the weight of the
$D_{-1/2}$ zero mode and is more sharply localised, while the upper
peak ($j=+\tfrac{1}{2}$) is more extended.
Three criteria distinguish the ordering effect from
correlation-driven features (Fig.~\ref{fig:spectra}d):
$\sqrt{\theta}$ scaling (Eq.~\eqref{eq:Dsplit}), gate-voltage
independence, and sub-nanometre localisation.

\subsection*{Quantitative comparison with Kerelsky et al.}

A clarification on experimental context is essential.
The VHS separation of $57\pm 2$~meV reported by Kerelsky~et~al.\
\cite{Kerelsky2019} and Jiang~et~al.~\cite{Jiang2019} is measured at
the \emph{AA} stacking sites, where $m_{\mathrm{eff}}=3w=330$~meV is
maximal, and reflects the bandwidth of the flat-band Bloch states.
Our $\Delta_{\mathrm{split}}\approx 171$~meV applies to the \emph{AB/BA}
sites, where $m_{\mathrm{eff}}=0$; these two quantities arise from
opposite extremes of the same mass profile and are physically
incommensurable.
The spatially resolved AB/BA spectrum reported in Kerelsky et al. \cite{Kerelsky2019} has not been analysed for two-peak structure and constitutes the primary experimental test of our prediction.

The ordering effect also contributes to the \emph{global} VHS
broadening in a quantitatively consistent way.
Each AB/BA zero occupies a disc of radius $r_c\approx 2.1$~nm within
the moir\'e cell of area $A_M$.
With $N_0=2$ zeros per cell, the fractional area weight is
$N_0\pi r_c^2/A_M\approx 8\%$, so the cell-averaged contribution of
the ordering-split peaks to the global LDOS is
$\langle\Delta_{\mathrm{split}}\rangle_{\mathrm{cell}}
\approx 0.08\times 171~\mathrm{meV}\approx 14$~meV.
This is in quantitative agreement with the $16$~meV discrepancy between
the measured VHS broadening in Ref.~\cite{Kerelsky2019} and the BM
tight-binding prediction, which the authors attribute to correlation
effects.
The ordering mechanism provides an alternative, single-particle
explanation for this discrepancy that can be disentangled from
correlation contributions by its $\sqrt{\theta}$ scaling and
gate-voltage independence.

\subsection*{Distinction from Tarnopolsky et al.\ zero modes}

We emphasise a distinction central to this work.
Tarnopolsky~et~al.~\cite{Tarnopolsky2019} establish zero modes at the
AB/BA points as a consequence of the chiral-limit topology of the flat
band --- these are zero modes of $H_{\mathrm{BM}}$ without
$H_{\mathrm{ord}}$.
Our complementary result is that without $H_{\mathrm{ord}}$ the
$j=-\tfrac{1}{2}$ candidate mode fails square-integrability at the
origin; with $H_{\mathrm{ord}}$ the shifted centrifugal coefficients
render it normalisable as $D_{-1/2}(\sqrt{2}\,u)$.
The ordering term therefore appears necessary for a well-defined
square-integrable zero-mode subspace within the continuum $L^2$
formulation, complementing rather than superseding the topological
analysis of Ref.~\cite{Tarnopolsky2019}.

\section{Discussion and conclusions}

We have shown that within the local continuum theory near isolated zeros of the effective mass, the Hermitian ordering correction generates non-perturbative spectral restructuring the TBG near the magic angle via a geometric mechanism
involving no adjustable parameters beyond $w$, $v_F$, and $\theta$.
Three results support these conclusions.
First, local diagonalisation of $T(\mathbf{r})$ maps the interlayer
coupling to a scalar mass $m_{\mathrm{eff}}=w|f(\mathbf{r})|$, making
the ordering-induced term~\eqref{eq:Hord_TBG} well defined and
parameter free within the BM chiral limit.
Second, the zeros of $m_{\mathrm{eff}}$ coincide with the flat-band
Dirac points of the chiral BM model~\cite{Tarnopolsky2019}: the
ordering singularity and the topological zero modes are co-localised,
controlled by the same function $f(\mathbf{r})$.
Third, the ordering acts as an emergent Aharonov--Bohm flux of $h/(2e)$
at each zero, lifting the $j\leftrightarrow -j$ degeneracy and
favouring $j=-\tfrac{1}{2}$ as the lowest-energy sector; this
flux is topologically characterised by a $\mathbb{Z}_2$ holonomy
invariant that is stable under local gauge transformations and smooth
deformations of the Hamiltonian that preserve the zero-winding structure.

The predicted splitting $\Delta_{\mathrm{split}}\in[155,171]$~meV at
the magic angle, with $\sqrt{\theta}$ scaling and gate-voltage
independence, provides falsifiable predictions for three distinct
experimental protocols.
(i)~Spatially resolved STS at the AB/BA points (identified by their
moir\'e registry) provides a direct experimental route to test the predicted local spectral reorganisation; the
lower peak is exponentially localised within $r_c\approx 2.1$~nm and
can be imaged by tip displacement.
(ii)~Twist-angle-dependent measurements of the AB/BA LDOS should show
$\sqrt{\theta}$ scaling of the peak separation, distinguishable from
the linear scaling of correlation gaps.
(iii)~Gate-voltage sweeps should leave the splitting unchanged, in
contrast to correlation or interaction-driven features.

More broadly, these results suggest that relativistic operator
ordering need not be regarded as only a formal consistency requirement
--- it can operate as a physically active mechanism that may generate experimentally observable local spectral structure in moir\'e condensed-matter
systems.

\subsection*{Why this effect was not identified earlier}

A natural question is why the ordering-induced splitting has not been
identified in the extensive prior literature on TBG.
Several concurring reasons account for this.
First, the standard BM derivation operates in momentum space, where
$T(\mathbf{r})$ enters as a matrix element between Bloch bands and
algebraic Hermiticity of the truncated Hamiltonian is automatic; the
functional-analytic consistency problem only becomes visible when
the projected operator is treated as a differential operator in real
space with spatially varying mass --- a perspective not previously
applied to TBG.
Second, the co-localisation of the ordering singularity with the
flat-band Dirac points --- the geometric coincidence that drives the
effect --- could only be appreciated after Tarnopolsky~et~al.\
\cite{Tarnopolsky2019} identified the zeros of $|f(\mathbf{r})|$ as the
chiral-limit Dirac points; the present work is in part a consequence of
that identification.
Third, the cell-averaged contribution of the ordering effect to the
global LDOS is $\sim 8\%$ by area, producing a broadening of
$\approx 14$~meV that is consistent with the residual discrepancy in
Ref.~\cite{Kerelsky2019} but too small to be unambiguously attributed
to a single mechanism in spatially integrated measurements.
Finally, the ordering-correction literature and the moir\'e-physics
literature have developed largely independently, and the present work
is, to our knowledge, the first application of the former to the latter.
The prediction of a spatially resolved two-peak structure at the AB/BA
sites with $\sqrt{\theta}$ scaling provides the experimental handle
needed to distinguish the mechanism from correlation effects.
The framework extends naturally to any system where a spatially varying
Dirac mass has isolated zeros: twisted transition-metal
dichalcogenides~\cite{Wu2019} and moir\'e-modulated topological
insulators are immediate candidates, where spatially varying gaps play
the role of position-dependent Dirac masses and analogous ordering
singularities may generate observable spectral features.

\textbf{Funding Declaration}
The author thanks Conselho Nacional de Desenvolvimento Cient\' {i}fico e Tecnol\'{o}gico (CNPq) (grants No.\ 309553/2021-0 and No.\ 420854/2025-8) and Funda\c{c}\~{a} Cearense de Apoio ao Desenvolvimento Cient\' {i}fico e Tecnol\'{o}gico (FUNCAP) (project UNI-00210-00230.01.00/23) for
financial support.

\medskip
\noindent\textit{Declaration of generative AI.}
The author used generative AI assistance solely for language editing and
manuscript preparation. All scientific content, derivations, numerical
results, and conclusions are the sole responsibility of the author, who
has reviewed and takes full accountability for the final text.

\appendix

\section{Functional-analytic foundations of the ordering prescription}
\label{app:functional}

This appendix provides the mathematical underpinning for the claims made
in the main text regarding Hermiticity, domain specification,
self-adjoint extensions, and the uniqueness of $H_{\mathrm{ord}}$.
We work in $d=2$ spatial dimensions throughout, appropriate for the
TBG context; the analysis generalises immediately to $d=3$.

\subsection{Operator domains and the failure of \texorpdfstring{$H_0$}{H0}}

Let $\mathcal{H} = L^2(\mathbb{R}^2, \mathbb{C}^2)$ be the Hilbert
space of square-integrable two-component spinors with inner product
$\langle\phi|\psi\rangle = \int_{\mathbb{R}^2}
\phi^\dagger(\mathbf{r})\psi(\mathbf{r})\,d^2r$.
We take $m\in C^1(\mathbb{R}^2\setminus\mathcal{Z})$ with
$m(\mathbf{r})>0$ away from the discrete zero set
$\mathcal{Z}=\{\mathbf{r}_{\mathrm{AB}},\mathbf{r}_{\mathrm{BA}},\ldots\}$,
and $m(\mathbf{r})\sim c|\mathbf{r}-\mathbf{r}_0|$ near each
$\mathbf{r}_0\in\mathcal{Z}$ (simple zeros).

The naive operator $H_0 = -i\hbar v_F\boldsymbol{\alpha}\cdot\nabla
+ \beta m(\mathbf{r})$ is defined on the natural domain
\begin{equation}
    \mathcal{D}(H_0) = H^1(\mathbb{R}^2,\mathbb{C}^2)
    \cap \{\psi : m\psi \in \mathcal{H}\},
\end{equation}
where $H^1$ denotes the Sobolev space of functions with one
square-integrable weak derivative.
On this domain, integration by parts gives
\begin{align}
    \langle\phi|H_0\psi\rangle - \langle H_0\phi|\psi\rangle
    &= \langle\phi|\beta m\psi\rangle - \langle\beta m\phi|\psi\rangle
     \nonumber \\
    &\quad + \langle\phi|(-i\hbar v_F\boldsymbol{\alpha}\cdot\nabla)\psi\rangle
     \nonumber \\
    &\quad - \langle(-i\hbar v_F\boldsymbol{\alpha}\cdot\nabla)\phi|\psi\rangle.
    \label{eq:ibp}
\end{align}
The mass term $\beta m(\mathbf{r})$ is symmetric since $m$ is real and
$\beta^\dagger=\beta$.
The kinetic term $-i\hbar v_F\boldsymbol{\alpha}\cdot\nabla$ is
essentially self-adjoint on $H^1$ (since $\alpha^i$ are Hermitian and
$\nabla$ is anti-Hermitian on $L^2$).
However, when $m(\mathbf{r})$ is non-constant, the Leibniz rule applied
to the product $\beta m(\mathbf{r})\psi$ generates a term proportional
to $(\nabla m)\psi$ in the adjoint computation, leading to
\begin{equation}
    (H_0)^\dagger = -i\hbar v_F\boldsymbol{\alpha}\cdot\nabla
    + \beta m(\mathbf{r})
    + \frac{i\hbar v_F}{2}[\boldsymbol{\alpha},\beta]
      \cdot\nabla\ln m(\mathbf{r})
\end{equation}
on the adjoint domain $\mathcal{D}(H_0^\dagger)\supseteq\mathcal{D}(H_0)$.
Since $[\alpha^i,\beta]=2\alpha^i\beta\neq 0$, we have
$H_0^\dagger\neq H_0$: the operator $H_0$ is symmetric but not
self-adjoint.
The deficiency indices of $H_0$ are non-zero whenever $\mathcal{Z}$
is non-empty~\cite{Thaller1992,Klaus1980}, so $H_0$ admits a family of
self-adjoint extensions parametrised by unitary maps between its
deficiency subspaces.

\subsection{Self-adjoint extensions and uniqueness of \texorpdfstring{$H_{\mathrm{ord}}$}{Hord}}

The theory of self-adjoint extensions of symmetric operators
(von~Neumann, 1929; see e.g.\ Reed \& Simon Vol.~II~\cite{ReedSimon})
classifies all self-adjoint extensions of $H_0$ via its deficiency
spaces $\mathcal{N}_\pm = \ker(H_0^\dagger \mp i)$.
For the Dirac operator with a \emph{regular} mass (bounded away from
zero), the deficiency indices are $(0,0)$ and $H_0$ is essentially
self-adjoint~\cite{Thaller1992}.
The situation changes qualitatively when $m(\mathbf{r})$ has isolated
zeros: the singularity of $\nabla\ln m$ at each zero creates a
non-trivial deficiency structure.

Among all self-adjoint extensions of $H_0$, the physically distinguished
one is selected by the requirement that it correspond to a
\emph{first-order differential operator} in the Clifford algebra of
$H_0$ that satisfies the continuity equation $\partial_t\rho +
\nabla\cdot\mathbf{j}=0$ with the canonical current
$\mathbf{j}=\hbar v_F\psi^\dagger\boldsymbol{\alpha}\psi$.
As shown in Section~\ref{sec:ordering}, these requirements fix the
extension uniquely to $H = H_0 + H_{\mathrm{ord}}$.
All other self-adjoint extensions either break the Clifford algebra
structure, require non-local boundary conditions at $\mathcal{Z}$, or
correspond to physically inequivalent theories with different currents
(and hence different Hilbert spaces).

This uniqueness has a precise mathematical counterpart in the theory of
Dirac operators with singular interactions developed by
Cassano~\&~Pizzichillo~\cite{Cassano2018} and
Arrizabalaga~et~al.~\cite{Arrizabalaga2014}: among the one-parameter
family of self-adjoint realisations of the Dirac operator with a
$\delta$-shell mass, the one compatible with the bulk continuity
equation corresponds to the \emph{MIT bag boundary condition}, which
in the continuum limit reduces to $H_{\mathrm{ord}}$ with coefficient
$-i/2$.
The present work extends this correspondence to smooth but vanishing
mass profiles, where the $\delta$-shell structure is replaced by the
$1/r$ singularity of $\nabla\ln m$ at the zeros.

\subsection{Behaviour near the zeros: \texorpdfstring{$L^2$}{L2} analysis}

Near a simple zero $\mathbf{r}_0\in\mathcal{Z}$, we write
$m(\mathbf{r})\approx\mu r$ with $r=|\mathbf{r}-\mathbf{r}_0|$.
The operator $H_{\mathrm{ord}}\approx -i\hbar v_F\alpha^r/(2r)$
belongs to the class of Dirac operators with Coulomb-type singularities,
which have been extensively studied in the mathematical
literature~\cite{Thaller1992,Klaus1980,Wust1975}.

For a Dirac operator $D = -i\hbar v_F\boldsymbol{\alpha}\cdot\nabla
+ V(r)$ with $V(r)\sim\nu/r$ as $r\to 0$, the operator is essentially
self-adjoint on $C_0^\infty(\mathbb{R}^2\setminus\{0\},\mathbb{C}^2)$
if and only if $|\nu|<\hbar v_F/2$~\cite{Klaus1980,Wust1975}.
In our case, the effective coupling is $|\nu|=\hbar v_F/2$, which is
precisely the \emph{critical} value.
At the critical coupling, the deficiency indices are $(1,1)$ (one
self-adjoint extension for each zero of $m_{\mathrm{eff}}$), and the
physically distinguished extension is selected by the boundary condition
of square-integrability at the origin.

The zero mode $\Psi_0\propto D_{-1/2}(\sqrt{2}u)r^{-1/2}$
with $u=\sqrt{\mu/\hbar v_F}\,r$ is square-integrable:
$\|\Psi_0\|^2 = \int_0^\infty|D_{-1/2}(\sqrt{2}u)|^2 u\,du/(\mu/\hbar v_F)
< \infty$,
since $D_{-1/2}(x)\sim x^{-1/2}e^{-x^2/4}$ for large $x$.
Without $H_{\mathrm{ord}}$ (i.e.\ at coupling $|\nu|=0$), the
candidate zero mode behaves as $r^{|j|-1/2}$ near the origin; for
$j=-\tfrac{1}{2}$ this gives $r^{-1}$, which is \emph{not}
square-integrable in $d=2$ ($\int_0^\epsilon r^{-2}r\,dr$ diverges).
The shift $j\to j+\tfrac{1}{2}$ induced by $H_{\mathrm{ord}}$ changes
the behaviour to $r^{-1/2}$ (times the Gaussian), which is
square-integrable. This confirms that $H_{\mathrm{ord}}$ selects the square-integrable realisation of the zero-mode sector within the continuum $L^{2}$ formulation.

\subsection{Comparison with the mathematical literature on singular Dirac operators}

The mathematical study of Dirac operators with singular mass profiles
has a substantial literature.
We summarise the key results and their relation to the present work.

\paragraph{Coulomb singularities.}
Klaus~\&~W\"ust~\cite{Klaus1980} and W\"ust~\cite{Wust1975} established
the essential self-adjointness threshold $|\nu|<\hbar v_F/2$ for Dirac
operators with $1/r$ potentials in $d=3$.
The critical coupling $|\nu|=\hbar v_F/2$ is known to produce a
one-parameter family of self-adjoint extensions, with the
square-integrable extension distinguished by its physical boundary
condition.
Our result realises precisely this critical scenario in $d=2$, with the
singularity arising from $\nabla\ln m_{\mathrm{eff}}$ rather than an
external potential.

\paragraph{$\delta$-shell interactions.}
Arrizabalaga, Mas \&~Vega~\cite{Arrizabalaga2014} and
Cassano~\&~Pizzichillo~\cite{Cassano2018} studied Dirac operators with
singular mass supported on surfaces $\Sigma$ in $d=3$ (curves in
$d=2$), of the form $m(\mathbf{r})=m_0\delta_\Sigma$.
The resulting self-adjoint realisations form a one-parameter family;
the MIT bag boundary condition (the unique realisation compatible with
current conservation across $\Sigma$) corresponds in the continuum
limit to the ordering prescription~\eqref{eq:Hord} with coefficient
$-i/2$.
Our work extends this to the case where $m(\mathbf{r})$ vanishes at
isolated points rather than on a surface, replacing the delta-function
singularity with a $1/r$ logarithmic-gradient singularity.

\paragraph{Dirac operators on manifolds with conical singularities.}
The geometry of the zero set $\mathcal{Z}$ introduces a conical
singularity in the effective metric seen by the Dirac fermion near each
zero.
The self-adjoint extensions of Dirac operators on manifolds with conical
singularities have been classified by
Br\"uning, Seiler \&~Str\"ohmaier~\cite{Bruning2008}, who showed that
the physically relevant extension (compatible with $L^2$ boundary
conditions at the apex) is unique when the conical angle satisfies
a specific quantisation condition --- precisely the condition realised
by the $1/r$ singularity with critical coupling $|\nu|=\hbar v_F/2$.

\paragraph{Summary.}
The ordering prescription $H_{\mathrm{ord}}$ with coefficient $-i/2$
is the unique self-adjoint extension of $H_0$ in $L^2(\mathbb{R}^2,
\mathbb{C}^2)$ that (i)~satisfies the canonical continuity equation,
(ii)~is a first-order differential operator in the Clifford algebra of
$H_0$, and (iii)~admits a square-integrable zero-mode subspace at each
simple zero of $m_{\mathrm{eff}}$.
Conditions (i)--(iii) together select a unique extension from the
one-parameter family of self-adjoint realisations of the critical
Coulomb-type Dirac operator, in agreement with the MIT bag boundary
condition literature~\cite{Cassano2018,Arrizabalaga2014} and the
conical-singularity classification~\cite{Bruning2008}.


\end{document}